\documentclass{PoS}
\usepackage{subfigure}

\usepackage{graphicx}
\usepackage{grffile}

\title{Search for electroweak supersymmetric particle production in final states with two leptons and missing transverse momentum with the ATLAS detector}

\ShortTitle{Search for electroweak supersymmetric particles}

\author{\speaker{Janet DIETRICH}\thanks{On behalf of the ATLAS collaboration}\\
         Deutsches Elektronen-Synchrotron\\
        E-mail: \email{janet.dietrich@cern.ch}}

\abstract{

Searches for the production of electroweak supersymmetric particles decaying into final states with exactly two isolated, oppositely-charged leptons
(electrons, muons), no reconstructed jets and missing transverse momentum are performed using \datalumi\ of 2012 proton-proton collision data at $\sqrt{s}=8\TeV$ recorded with the general purpose detector ATLAS at the Large Hardon
Collider. In the absence of any significant excess with respect to the prediction from Standard Model processes, the results are interpreted in the framework of simplified Supersymmetry
models~\cite{2LDGCONF2013}.

}

\FullConference{The European Physical Society Conference on High Energy Physics -EPS-HEP2013\\
		18-24 July 2013\\
		Stockholm, Sweden}
\def\TeV{\ifmmode {\mathrm{\ Te\kern -0.1em V}}\else
                   \textrm{Te\kern -0.1em V}\fi}%
\def\GeV{\ifmmode {\mathrm{\ Ge\kern -0.1em V}}\else
                   \textrm{Ge\kern -0.1em V}\fi}%

\def\ifb{\mbox{fb$^{-1}$}}
\def\susy#1{\ensuremath{\tilde{#1}}}%
\def\slepton{\ensuremath{\tilde{\ell}}}
\def\chinoonep{\ensuremath{\mathchoice%
      {\displaystyle\raise.4ex\hbox{$\displaystyle\tilde\chi^+_1$}}%
         {\textstyle\raise.4ex\hbox{$\textstyle\tilde\chi^+_1$}}%
       {\scriptstyle\raise.3ex\hbox{$\scriptstyle\tilde\chi^+_1$}}%
 {\scriptscriptstyle\raise.3ex\hbox{$\scriptscriptstyle\tilde\chi^+_1$}}}}
\def\chinoonepm{\ensuremath{\mathchoice%
      {\displaystyle\raise.4ex\hbox{$\displaystyle\tilde\chi^\pm_1$}}%
         {\textstyle\raise.4ex\hbox{$\textstyle\tilde\chi^\pm_1$}}%
       {\scriptstyle\raise.3ex\hbox{$\scriptstyle\tilde\chi^\pm_1$}}%
 {\scriptscriptstyle\raise.3ex\hbox{$\scriptscriptstyle\tilde\chi^\pm_1$}}}}
 \def\chinoonemp{\ensuremath{\mathchoice%
      {\displaystyle\raise.4ex\hbox{$\displaystyle\tilde\chi^\mp_1$}}%
         {\textstyle\raise.4ex\hbox{$\textstyle\tilde\chi^\mp_1$}}%
       {\scriptstyle\raise.3ex\hbox{$\scriptstyle\tilde\chi^\mp_1$}}%
 {\scriptscriptstyle\raise.3ex\hbox{$\scriptscriptstyle\tilde\chi^\mp_1$}}}}
\def\ninotwo{\ensuremath{\mathchoice%
      {\displaystyle\raise.4ex\hbox{$\displaystyle\tilde\chi^0_2$}}%
         {\textstyle\raise.4ex\hbox{$\textstyle\tilde\chi^0_2$}}%
       {\scriptstyle\raise.3ex\hbox{$\scriptstyle\tilde\chi^0_2$}}%
 {\scriptscriptstyle\raise.3ex\hbox{$\scriptscriptstyle\tilde\chi^0_2$}}}}

\newcommand{\datalumi}{20.3\,\ifb}
\newcommand{\exclMslepMinZero}{90\,GeV}
\newcommand{\exclMslepMaxZero}{320\,GeV}

\newcommand{\exclMconeMinTwenty}{130\,GeV}
\newcommand{\exclMconeMaxTwenty}{450\,GeV}
\newcommand{\gaugino}[2]{\ensuremath{\tilde\chi_{#1}^{#2}}}
\newcommand{\neutralino}[1]{\gaugino{#1}{0}}
\newcommand{\chargino}[1]{\gaugino{#1}{\pm}}
\newcommand{\sneutrino}{\ensuremath{\tilde\nu}}
\newcommand{\SRWW}{SR-WW} 
\newcommand{\SRWWa}{SR-WWa} 
\newcommand{\SRWWb}{SR-WWb} 
\newcommand{\SRWWc}{SR-WWc} 

\newcommand{\SRmtt}{SR-\mtt}
\newcommand{\SRmtta}{SR-\mtta} 
\newcommand{\SRmttb}{SR-\mttb} 

\newcommand{\mtt}{\ensuremath{m_{\mathrm{T2}}}} 
\newcommand{\mtta}{\ensuremath{m_{\mathrm{T2},\,90}}} 
\newcommand{\mttb}{\ensuremath{m_{\mathrm{T2},\,120}}} 
 
\newcommand{\mll}{\ensuremath{m_{\ell\ell}}}
\newcommand{\mttwo}{\ensuremath{m_\mathrm{T2}}}
\def\ee{\ensuremath{e^+ e^-}}%
\def\mumu{\ensuremath{\mathrm{\mu^+ \mu^-}}}%

\newcommand{\emu}{$e^\pm\mu^\mp$}
\def\pT{\ensuremath{p_{\mathrm{T}}}} 

\newcommand{\pTll}{\ensuremath{p_\mathrm{T,\ell\ell}}}
\newcommand{\dphill}{\ensuremath{\Delta\phi_{\ell\ell}}}
\newcommand{\METrel}{\ensuremath{E_\mathrm{T}^\mathrm{miss,rel}}}

\begin{document}

\section{Introduction}

Weak scale Supersymmetry (SUSY) is one of the best motivated extensions of the Standard Model (SM), providing a possible solution to the hierarchy problem and a viable dark matter candidate in the form of the lightest supersymmetric particle (LSP). The dominant
SUSY production channels at the LHC depend on the masses of the sparticles. In scenarios where the masses of the first and second generation sfermions and gluinos are larger than few TeVs,
direct production of weak gauginos (charginos, $\susy{\chi}^{\pm}$ and neutralinos, $\susy{\chi}^{0}$) as well as sleptons ($\susy{\ell}$ and $\susy{\nu}$) 
may be the dominant SUSY process. The searches presented here target final states with two leptons and missing energy. They can be produced by \slepton\slepton\ production followed by $\slepton^\pm \to \ell^\pm \neutralino{1}$ decay, or by 
\chinoonepm\chinoonemp\ production followed by $\chargino{1} \to (\slepton^\pm\nu$ or $\ell^\pm\sneutrino) \to \ell^\pm\nu\neutralino{1}$ decay with two additional neutrinos contributing to
the missing transverse momentum. If the lightest chargino is heavier
than the LSP, the chargino decays as $\chargino{1}\to W^\pm\neutralino{1}$, producing an on- or off-shell $W$ boson.


\section{Event selection}

Based on the target signal model, five signal regions (SRs) are designed (see Table~\ref{SR}) selecting final states with two isolated opposite-sign leptons (electrons or muons), missing transverse momentum and
no jets (including b-tagged jets). \SRmtt\ signal regions are optimised to provide sensitivity to sleptons either through direct production or in chargino decays, while the \SRWW\ signal
regions are targeting chargino- and neutralino-pair production followed by on-shell $W$ decays.


\begin{table}[!htp]
\small
\begin{center}
\caption{\label{SR} Event selection criteria for the five signal regions with variables as defined in~\cite{2LDGCONF2013}}
\begin{tabular}{c|c|c|c|c|c}
\hline
                  & \SRmtta  & \SRmttb   & \SRWWa & \SRWWb & \SRWWc \\
\hline 
 lepton flavour   & \multicolumn{2}{c|}{\ee, \mumu, \emu} & \multicolumn{3}{c}{\emu} \\
 $\pT^{\ell1}$    & \multicolumn{2}{c|}{---} & \multicolumn{3}{c}{$>35\GeV$} \\
 $\pT^{\ell2}$    & \multicolumn{2}{c|}{---} & \multicolumn{3}{c}{$>20\GeV$} \\
 \mll             & \multicolumn{2}{c|}{$|\mll-m_Z|>10\GeV$} & $<80\GeV$ & $<130\GeV$ &     ---    \\
 \pTll            & \multicolumn{2}{c|}{---} & $>70\GeV$ & $<170\GeV$ & $<190\GeV$ \\
 \dphill          & \multicolumn{2}{c|}{---} & \multicolumn{3}{c}{$<1.8$~rad} \\
 \METrel          & \multicolumn{2}{c|}{$>40\GeV$} & $>70\GeV$ & \multicolumn{2}{c}{---} \\
 \mttwo           & $>90\GeV$ & $>110\GeV$ &    ---    & $>90\GeV$  & $>100\GeV$ \\
\hline

\end{tabular}
\end{center}
\end{table}

\section{Background estimation}

The main SM background comes from $WW$ diboson and top-pair production where two leptonically decaying $W$ bosons result in the same final state as the SUSY signal. Another significant source of background in the
same-flavour channel is $WZ$ and $ZZ$ production. These events are estimated by defining dedicated control regions for each background and extracting a normalization factor to be applied to the simulations in the signal regions. 
Leptons originating from heavy-flavour decays or photon conversion or mistakenly reconstructed hadronic jets can be mis-identified as signal leptons. This ``fake'' background is obtained in a fully data-driven way (matrix method). 
Other minor backgrounds such as $Z$\,+\,jets are estimated using the MC predictions.

\section{Results}
\begin{figure}[!ht]
  \centering
  \subfigure[]{
    \includegraphics[width=0.4\textwidth ]{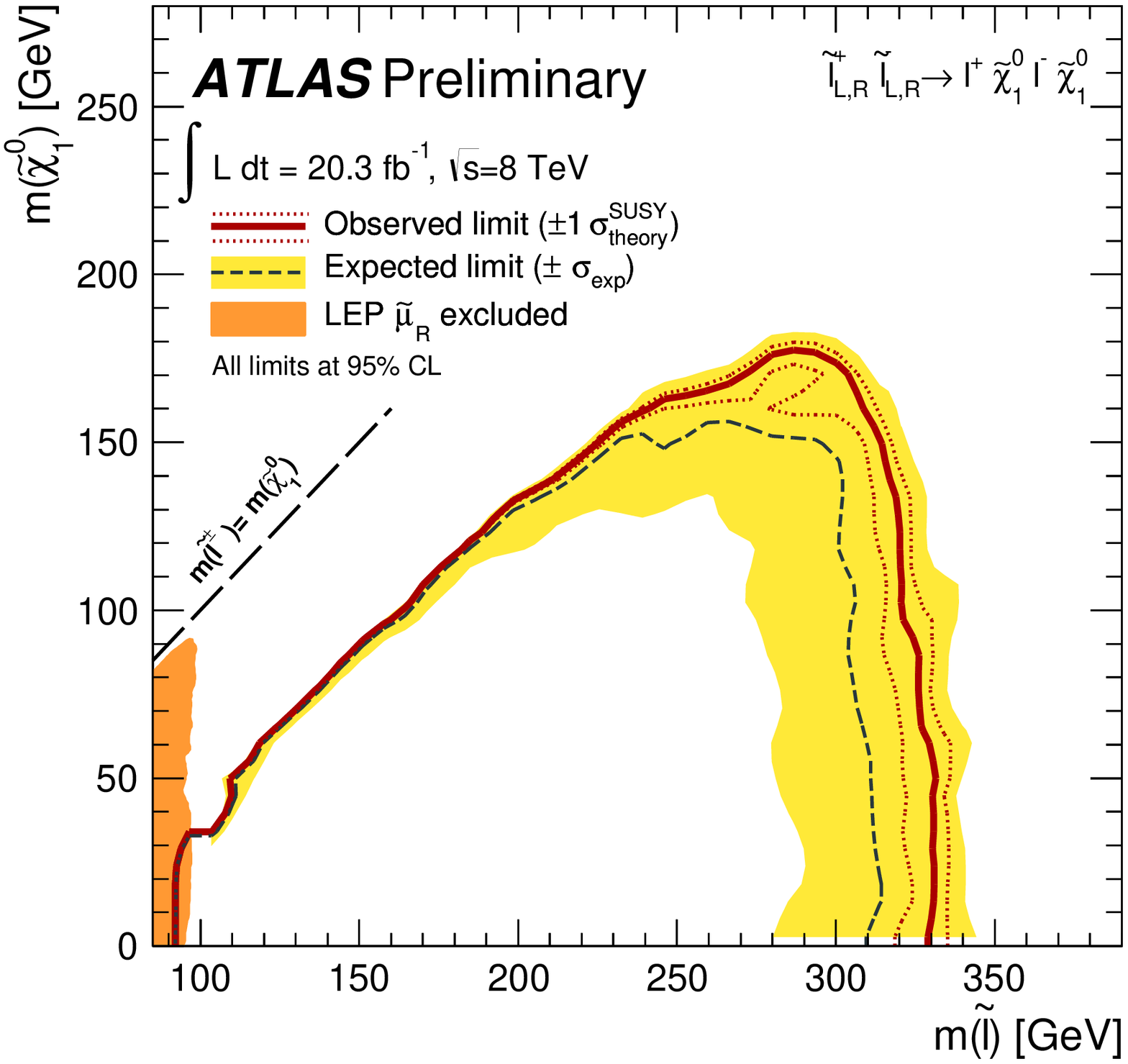}\label{fig_a}
     }
  \subfigure[]{
    \includegraphics[width=0.4\textwidth ]{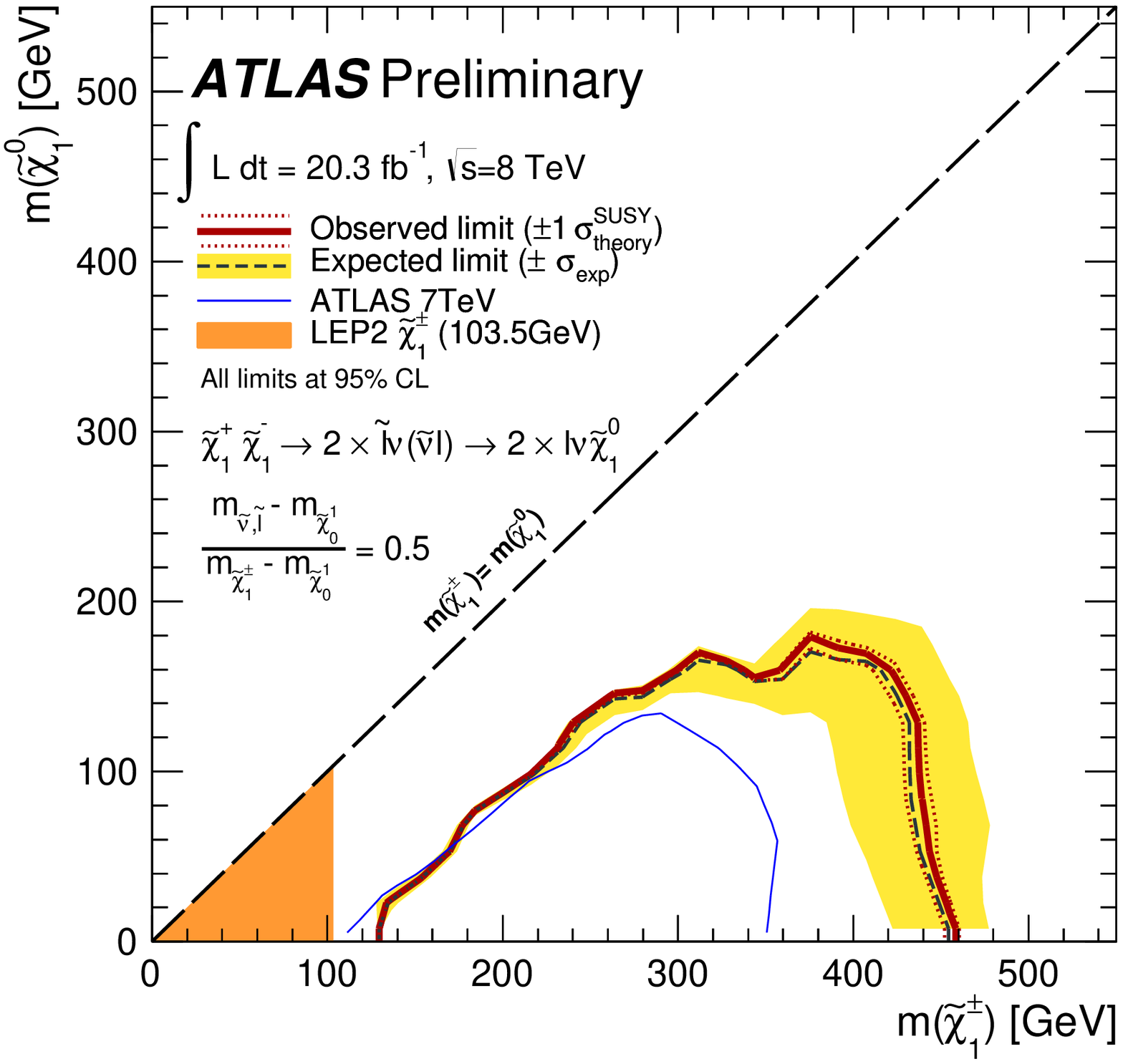}\label{fig_b}
     }

        \caption{\label{interpretation} \small
                95\% CL exclusion limits for both right- and left-handed (mass degenerate) selectron and smuon production in the
		$m_{\neutralino{1}}$--$m_{\slepton}$ plane (a) and 95\% CL exclusion limits for $\tilde{\chi}^\pm_1\tilde{\chi}^\mp_1$ pair production in the
		simplified model with sleptons and sneutrinos with $m_{\slepton}=m_{\sneutrino}=(m_{\chargino{1}}+m_{\neutralino{1}})/2$ (b)\cite{2LDGCONF2013}. The dashed and solid lines show the 95\% CL$_\textup{s}$ expected and observed limits, respectively, including all uncertainties except for the theoretical signal
                cross-section uncertainty (PDF and scale). The solid band around the expected limit shows the $\pm1\sigma$ result where
                all uncertainties, except those on the signal cross-sections, are considered. The $\pm1 \sigma$ lines around the observed limit represent the results obtained
                when moving the nominal signal cross-section up or down by the $\pm 1 \sigma$ theoretical uncertainty.
                Illustrated also are the LEP limits~\cite{lepsusy} on the mass of the right-handed smuon $\tilde{\mu}_{R}$ in (a) and on the mass of the chargino in (b).
                The blue line indicates the limit from the previous analysis with the 7\,TeV data~\cite{Aad:2012pxa}.}

\end{figure}

No significant excesses over the Standard Model predictions are observed.
In scenarios with direct sleptons decays, a common value for left- and right-handed slepton masses between \exclMslepMinZero\ and \exclMslepMaxZero\ is excluded at 95\% confidence level for a
massless neutralino (Figure~\ref{interpretation} a). In the scenario
of chargino-pair production, with wino-like charginos decaying into the lightest neutralino via an intermediate slepton, chargino masses between \exclMconeMinTwenty\ and \exclMconeMaxTwenty\
are excluded at 95\% confidence level for a $20\GeV$ neutralino (Figure~\ref{interpretation} b). In the scenario of chargino-pair production with $W$ boson decays, the excluded cross-section is above the model cross-section by a factor 1.9--2.8 in the \chargino{1} mass range 100--$190\GeV$ and then increases gradually to 4.7 when
reaching a \chargino{1} mass of $250\GeV$. The best sensitivity is obtained for the $(m_{\chargino{1}},m_{\neutralino{1}})=(100,0)\GeV$ mass point where $\sigma/\sigma_\mathrm{SUSY}=1.8$.


\begin{thebibliography}{99}

 



\bibitem{2LDGCONF2013}
ATLAS Collaboration, ATLAS-CONF-2013-049, http://cdsweb.cern.ch/record/1557779

\bibitem{lepsusy}

LEPSUSYWG, ALEPH, DELPHI, L3 and OPAL experiments, note LEPSUSYWG/01-03.1, 04-01.1, http://lepsusy.web.cern.ch/lepsusy/Welcome.html
\bibitem{Aad:2012pxa}
ATLAS Collaboration, \emph{Phys. Lett.} {\bf B718} (2013) 879-901 [{\tt hep-ex/1208.2884}]


\end{thebibliography}
\end{document}